\documentclass[12pt,a4paper]{article}
\usepackage{amsmath,amssymb,bm,ascmac,bbm,mathtools}
\usepackage[dvipdfmx, usenames]{color}
\usepackage[dvipdfmx]{graphicx}
\usepackage{xcolor}
\usepackage{here}
\usepackage{authblk}
\usepackage[hang,small,bf]{caption}
\usepackage[subrefformat=parens]{subcaption}
\usepackage{dcolumn}

\newcommand{\tdif}[2]{\frac{d#1}{d#2}}

\newcommand{\notsubset}{\not\subset}

\newcommand{\tr}{\text{tr}}

\newcommand{\mbb}{\mathbb}
\newcommand{\mfrak}{\mathfrak}

\newcommand{\ra}{\rangle}

\newcommand{\rank}{\text{rank}}

\newcommand{\mc}{\mathcal}
\newcommand{\uv}{\text{UV}}
\newcommand{\ir}{\text{IR}}
\newcommand{\wf}[1]{\widehat{\mfrak{#1}}}
\newcommand{\vecG}{\text{Vec}}
\newcommand{\vect}{\text{Vect}_{\mbb C}}
\newcommand{\svec}{\text{sVec}}

\newcommand{\fib}{\text{Fib}}
\newcommand{\ising}{\text{Ising}}

\begin{document}
\title{Refined half-integer condition on RG flows}
\author{Ken KIKUCHI}
\affil{International Center for Theoretical Sciences, Tata Institute of Fundamental Research, Shivakote, Hesaraghatta Hobli, Bengaluru North 560089, India}
\date{}
\maketitle

\begin{abstract}
Renormalization group flows are constrained by symmetries. Traditionally, we have made the most of 't Hooft anomalies associated to the symmetries. The anomaly is mathematically part of the data for the monoidal structure on symmetry categories. The symmetry categories sometimes admit additional structures such as braiding. It was found that the additional structures give further constraints on renormalization group flows. One of these constraints is the half-integer condition. The condition claims the following. Braidings are characterized by conformal dimensions. A symmetry object $c$ in a braided symmetry category surviving all along the flow thus has two conformal dimensions, one in ultraviolet $h_c^\text{UV}$ and the other in infrared $h_c^\text{IR}$. In a renormalization group flow with a renormalization group defect, they add up to a half-integer $h_c^\text{UV}+h_c^\text{IR}\in\frac12\mathbb Z$. We find a necessary condition for the sum to be half-integer. We solve some flows with the refined half-integer condition.
\end{abstract}

\makeatletter
\renewcommand{\theequation}
{\arabic{section}.\arabic{equation}}
\@addtoreset{equation}{section}
\makeatother

\section{Introduction and summary}
Pick a two-dimensional conformal field theory (CFT). Deform the ultraviolet (UV)  theory with relevant operators. What is the infrared (IR) theory? The renormalization group (RG) flow is constrained by symmetries surviving the relevant deformation. Since symmetry operators are defined to be topological \cite{GKSW14}, unbroken symmetry objects remain in effective theories regardless of energy scales. Therefore, the effective theories (in particular the IR theory) must be compatible with the surviving symmetry. Traditionally, we have made the most of 't Hooft anomalies \cite{tH79} of surviving symmetries. If the surviving symmetry $\mc C$ has a nontrivial anomaly, candidate theories with different anomalies (especially the trivial theory) cannot be the IR theory. These theories are ruled out from the candidates of the IR theory by the 't Hooft anomaly matching. The anomaly is mathematically part of the data for the monoidal structure on the symmetry category $\mc C$. Given two symmetry objects $c_1,c_2\in\mc C$, one can multiply (usually called fusion) them to get a new symmetry object $c_1\otimes c_2\in\mc C$. (If $\mc C$ is group-like, the multiplication is just the group multiplication.) The fusion is required to be associative; given three symmetry objects $c_1,c_2,c_3\in\mc C$, we must have an isomorphism $\alpha_{1,2,3}:(c_1\otimes c_2)\otimes c_3\cong c_1\otimes(c_2\otimes c_3)$. The isomorphisms assemble to a natural isormorphism $\alpha_{-,-,-}:(-\otimes-)\otimes-\stackrel{\cong}\Rightarrow-\otimes(-\otimes-)$ called associator. The associator is part of the data of the monoidal structure of the monoidal category $\mc C$.  The associator is required to obey consistency conditions; given four symmetry objects $c_1,c_2,c_3,c_4\in\mc C$, there are two ways to send $((c_1\otimes c_2)\otimes c_3)\otimes c_4$ to $c_1\otimes(c_2\otimes(c_3\otimes c_4))$. The two ways should give the same result. The consistency condition is called the pentagon axiom.\footnote{For a physics-oriented review, see \cite{BT17}. For a standard textbook, see \cite{EGNO15}.} (If $\mc C$ is a group $G$, the associator reduces to a cohomology class of the group cohomology $\alpha\in H^3(G,U(1))$, and the pentagon axiom reduces to the $3$-cycle condition on $\alpha$.) Since RG flows are given by monoidal functors,\footnote{A monoidal functor $F:\mc C\to\mc D$ preserves fusion, i.e., $\forall c,c'\in\mc C,\ F(c\otimes c')\cong F(c)\otimes F(c')$. When $\mc C,\mc D$ are groups, this reduces to a group homomorphism.} the associators of the surviving symmetry categories are preserved all along the RG flows \cite{KKARG}. This is the mathematical reason why 't Hooft anomalies are preserved under RG flows.

The 't Hooft anomaly matching is so powerful that it has been playing a central role in constraining IR theories. However, the surviving symmetry categories sometimes have additional structures. For example, if one picks a rational conformal field theory (RCFT) as a UV theory, its symmetry category has additional braiding structure.\footnote{In fact, it was proven \cite{HL94,H05} that RCFTs have modular fusion categories (MFC) which have nondegenerate braiding structures.} The structure is a natural isomorphism; given two objects $c_1,c_2\in\mc C$, a braiding is an isomorphism $c_{c_1,c_2}:c_1\otimes c_2\cong c_2\otimes c_1$. They assemble to a natural isomorphism $c_{-,-}:-\otimes-\stackrel\cong\Rightarrow-\otimes-$. The natural isomorphism is subject to another consistency condition called the hexagon axiom. (Just as the pentagon axiom, it originates from the consistency of two ways to send $(c_1\otimes c_2)\otimes c_3$ to $c_2\otimes(c_3\otimes c_1)$, and two ways to send $c_1\otimes(c_2\otimes c_3)$ to $(c_3\otimes c_1)\otimes c_2$.) Since any objects of the full UV symmetry category $\mc C$ admit braiding, objects in the surviving symmetry category $\mc S_\uv$ admit braiding as well. Since the symmetry objects remain all along the RG flow, the effective theories have to be compatible with the surviving symmetry $\mc S_\uv$. We saw the monoidal structure of $\mc S_\uv$ constrains the IR theory. Then, a natural question is this: Does the additional structure (braiding) impose additional constraints on RG flows? It turned out the answer is yes; the braiding structure of surviving symmetry category can rule out\footnote{The braiding structures of surviving symmetry categories also give mathematical explanation when and why which symmetries emerge in IR \cite{KK21,KKSUSY,KK22II,KK22free}. We will employ the constraints below as well.} candidate IR theories which are allowed by the constraints from the monoidal structure alone, namely 't Hooft anomaly matching. For example, \cite{TS18} computed mixed anomaly in $PSU(N)\times\mbb Z/N\mbb Z$ symmetry of $SU(N)_k$ Wess-Zumino-Witten (WZW) model. They concluded the anomaly matching allows an RG flow $SU(N)_k\to SU(N)_{\gcd(N,k)}$. To be concrete, the monoidal structure alone cannot rule out $SU(3)_4\to SU(3)_1$. However, the flow does not satisfy additional constraints originating from the braiding structure \cite{KKWZW}. One of the additional constraints is the half-integer condition
\begin{equation}
    h^\uv_c+h^\ir_{F(c)}\in\frac12\mbb Z\quad(c\in\mc S_\uv),\label{halfintegercond}
\end{equation}
where $h_c$ is the conformal dimension of $c$. The condition was conjectured in \cite{KKWZW} and proven in \cite{KKWitt,KKhalfinteger} for RG flows with RG defects \cite{G12}.\footnote{It was called RG domain wall or RG interface in \cite{G12}. According to the terminology in \cite{CFLS19}, an interface, defect, and domain wall are defined as follows. One can smoothly vary a parameter in a theory (such as coupling constant). The resulting object is called smooth interface or simply interface. On the other hand, one can abruptly change the parameter. The resulting object lying where the parameter jumps is called a sharp interface or defect. Finally, co-dimension one dynamical excitations are called domain walls. As we will see below, RG domain wall or RG interface can be viewed as an interface lying where relevant coupling constants abruptly jump. Thus, following this terminology, it is more appropriate to call it an RG defect.} Since the half-integer condition is our main character, let us elaborate on its statement.

First, in a diagonal RCFTs, there is a one-to-one correspondence between primary operators and (zero-form) symmetry operators called Verlinde lines \cite{V88}.\footnote{While there is a general proof of the correspondence, we find it more educational to look at a simple example. Consider the Yang-Lee model or $M(5,2)$ in the notation of the yellow book \cite{FMS}. It is a diagonal RCFT with two primary operators with conformal dimensions $0,-\frac15$. We place the theory on a two-torus with modular parameter $\tau$. Its modular $S$-matrix is given by (in the basis)
\[ S=-\sqrt{\frac2{5-\sqrt5}}\begin{pmatrix}1&-\zeta^{-1}\\-\zeta^{-1}&-1\end{pmatrix}\quad(\zeta:=\frac{1+\sqrt5}2). \]
Since a symmetry operator is topological, it commutes with the energy-momentum tensor, and hence commutes with all Virasoro generators. Therefore, symmetry actions cannot change conformal dimensions. We learn (internal) symmetries act diagonally on the primaries. Let us denote a putative symmetry operator by $c$, and its action on the primaries as $a,b$, respectively. This should make it clear that the maximum number of (independent) symmetry transformations is two. In terms of partition function, if we insert the symmetry operator parallel to a `time' slice, by shifting and acting it on the Hilbert space $\mc H$, we get the twisted partition function
\[ Z^c(\tau,\bar\tau):=\tr_\mc H\left(\hat c q^{L_0-c/24}\bar q^{\bar L_0-c/24}\right)=a|\chi_0(\tau)|^2+b|\chi_{-\frac15}(\tau)|^2, \]
where $c=-\frac{22}5$. The RHS can be written as
\[ Z^c(\tau,\bar\tau)=\sum_{i,j}(M_c)_{i,j}\chi_i(\tau)\bar\chi_j(\bar\tau) \]
with the mass matrix
\[ M_c=\begin{pmatrix}a&0\\0&b\end{pmatrix}. \]
The actions $a,b$ cannot be arbitrary. It should obey the Cardy condition \cite{C86,PZ00}; since we can view the other direction as `time,' the twisted partition function should have a physical interpretation in terms of (defect) Hilbert space. In equations, we perform the modular $S$-transformation. The symmetry operator $c$ is now inserted along the `time' direction, and it intersects a `time' slice. Thus, the original Hilbert space is modified to the defect Hilbert space $\mc H_c$. For the twisted partition function
\begin{equation}
    Z_c(\tau,\bar\tau):=Z^c(-1/\tau,-1/\bar\tau)=\tr_{\mc H_c}\left(q^{L_0-c/24}\bar q^{\bar L_0-c/24}\right)=\sum_{i,j}(SM_cS^\dagger)_{i,j}\chi_i(\tau)\bar\chi_j(\bar\tau)\label{defectHilbertspace}
\end{equation}
to have physical interpretation, all the coefficients have to be natural numbers \begin{equation}
    \forall i,j,\quad(SM_cS^\dagger)_{i,j}\stackrel!\in\mbb N.\label{cardycond}
\end{equation}
With the modular $S$-matrix and our mass matrix, this condition reduces to components of
\[ \begin{pmatrix}A&B\\B&C\end{pmatrix} \]
with $A:=\frac{(5+\sqrt5)a+(5-\sqrt5)b}{10},B:=\frac{-a+b}{\sqrt5},C:=\frac{(5-\sqrt5)a+(5+\sqrt5)b}{10}$ be natural numbers. By adding $A$ and $C$, we get
\[ a+b,\frac{-a+b}{\sqrt5}\in\mbb N. \]
Clearly, the identity operator $a=1=b$ satisfies the Cardy condition. Since the maximum number of symmetry operators is two, and we have already found one, we only have to fine another. For the other solution to the Cardy condition to make the second ratio a natural number, the difference of $a$ and $b$ must be a multiple of $\sqrt5$. Since we know $\zeta=\frac{1+\sqrt5}2,-\zeta^{-1}=\frac{1-\sqrt5}2$, these numbers could work as $a$ and $b$. For the difference $-a+b$ to be positive, we have to take
\[ a=-\zeta^{-1},\quad b=\zeta. \]
Then, $a+b=1,\frac{-a+b}{\sqrt5}=1$, and the Cardy condition is satisfied. Since we have found two solutions, we can stop here. A general proof shows that a symmetry operator $c_i$ corresponding to a primary $\phi_i$ (with $\phi_1$ the identity operator) acts on a primary operator $\phi_j$ by
\begin{equation}
    c_i\phi_j=\frac{S_{ij}}{S_{1j}}\phi_j,\label{diagonalaction}
\end{equation}
where $S_{ij}$ is the modular $S$-matrix.} The braiding of Verlinde lines are characterized by the conformal dimensions of the corresponding primary operators.\footnote{For example, a double braiding of $c_i,c_j$ is given by
\[ c_{c_j,c_i}\circ c_{c_i,c_j}\cong\bigoplus_{k=1}^{\rank(\mc C)}{N_{ij}}^k\frac{e^{2\pi ih_k}}{e^{2\pi i(h_i+h_j)}}id_k, \]
where ${N_{ij}}^k\in\mbb N$ is the fusion coefficient counting the number of (simple) object $c_k$ in the fusion $c_i\otimes c_j$, and $id_k$ is the identity morphism at $c_k$. Here, $\rank(\mc C)$ is the rank of $\mc C$ counting the number of simple objects in $\mc C$.} Therefore, if a symmetry operator $c$ survives a relevant deformation, and if the RG flow ends up in another diagonal\footnote{This is an assumption. However, many non-diagonal RCFTs can be described in terms of diagonal ones. For example, the three-state Potts model is non-diagonal as a representation of the Virasoro algebra, but it is diagonal as a representation of the extended $W_3$ algebra. In addition, many non-diagonal RCFTs are obtained by gauging anomaly-free (discrete) symmetries in diagonal RCFTs. If one is interested in RG flows of such non-diagonal RCFTs, one can study the flow from those in diagonal cases when the RG flows are triggered by operators invariant under the gauged symmetries employing the commutativity of relevant deformation and discrete gauging \cite{KK22II}.} RCFT, it admits two conformal dimensions, one in UV $h_c^\uv$ and the other in IR $h_{F(c)}^\ir$, where $F:\mc S_\uv\to\mc S_\ir$ is the monoidal functor giving the RG flow. It was shown in \cite{KKWitt,KKhalfinteger} that in RG flows with RG defects, their conformal dimensions should add up to half-integers (\ref{halfintegercond}). (Thus, the monoidal functor is \textit{not} braided.) The RG defects \cite{G12} enter the proof as follows. As in our case, suppose a relevant deformation of a UV RCFT triggers an RG flow to an IR RCFT. By assumption, if we turn on the relevant coupling(s) in the whole two-dimensional space(time), the UV RCFT turns into the IR RCFT in the whole space(time). Now, let us turn on the relevant coupling(s) in only half of the space(time). Then, the other half remains the UV RCFT, while the deformed half flows to the IR RCFT. The system develops a defect between two theories.\footnote{This picture makes it clear that the interface can be viewed as an object inserted where the relevant couplings abruptly jump from zero to non-zero values. Therefore, in the terminology of \cite{CFLS19}, we believe it is appropriate to call it an RG defect rather than RG domain wall or RG interface.} The defect is called an RG defect (or RG domain wall/interface). Gaiotto constructed the defect for RG flows between diagonal cosets, however, he claimed the construction works more generally. In fact, the intuition does not seem to have any obstruction for such RG flows to admit RG defects. If an RG flow between two RCFTs admits an RG defect, we can employ the folding trick to place two RCFTs on the same half of the space(time). (An RG defect becomes an RG boundary condition.) Then, we can consider a product symmetry operator $c F(c)$. It has (chiral) disorder (a.k.a. twist) operator with conformal dimension $(h_c^\uv+h_{F(c)}^\ir,0),(0,h_c^\uv+h_{F(c)}^\ir)$ at the end as the product symmetry operator obeys the trivial fusion $c F(c)\otimes1\cong c F(c)$. Gaiotto argued \cite{G12} that the disorder operators define extended vertex operator (super)algebra. Therefore, the sum has to be a half-integer (\ref{halfintegercond}).\footnote{More precisely, the RG flow has to be simple. An RG flow is called simple \cite{KKWZW} if it cannot be divided into more than one RG flows. On the other hand, a sequence of more than one RG flows is called non-simple. An example of non-simple RG flow is tetracritical Ising model $M(6,5)\to$ tricritical Ising model $M(5,4)\to$ critical Ising model $M(4,3)$. In order to get a non-simple RG flow, one has to deform with relevant operators more than once because one relevant deformation brings a UV RCFT to an IR RCFT. In other words, non-simple RG flows in general do not support nontrivial surviving symmetries throughout the sequence of RG flows. Concretely, if the first RG flow preserves $\mc S_1$ and the second $\mc S_2$, $\mc S_1$ and $\mc S_2$ do not have to have common symmetry objects except the identity object. When one tries to solve an RG flow, one performs relevant deformation only once. Thus, the flows are believed to be simple.}

Before we explain our main result, refinement of the half-integer condition, we would like to briefly recall the definitions of vertex operator (super)algebras. A vertex operator algebra (VOA)\footnote{The VOAs were introduced in \cite{B85}. For standard textbooks, see, say, \cite{FLM88,LL04}.} is a quadruple $(V,Y,\mathbf1,\omega)$ of $\mbb Z$-graded vector space $V=\coprod_{n\in\mbb Z}V_{(n)}$, vertex operator map $Y(-,x):V\to(\text{End}V)[[x,x^{-1}]]$ sending $v\in V$ to $Y(v,x)=\sum_{n\in\mbb Z}v_nx^{-n-1}$, and two vectors $\mathbf1\in V$ called the vacuum vector and $\omega\in V_{(2)}$ called the conformal vector obeying six axioms: 1) truncation condition $\forall u,v\in V,\ Y(u,x)v\in V((x))$, 2) vacuum property $Y(\mathbf1,x)=1_V$, 3) creation property $Y(v,x)\mathbf1\in V[[x]],\lim_{x\to0}Y(v,x)\mathbf1=v$, 4) the Jacobi identity $x_0^{-1}\delta(\frac{x_1-x_2}{x_0})Y(u,x_1)Y(v,x_2)-x_0^{-1}\delta(\frac{x_2-x_1}{-x_0})Y(v,x_2)Y(u,x_1)=x_2^{-1}\delta(\frac{x_1-x_0}{x_2})Y(Y(u,x_0)v,x_2)$, 5) grading restriction $\forall n\in\mbb Z,\dim V_{(n)}<\infty$ and $V_{(n)}=0$ for sufficiently negative $n$, and 6) Virasoro algebra relations $[L(m),L(n)]=(m-n)L(m+n)+\frac1{12}(m^3-m)\delta_{m+n,0}c_V$ for $m,n\in\mbb Z$ with $Y(\omega,x)=\sum_{n\in\mathbb Z}L(n)x^{-n-2}$ and central charge $c_V\in\mbb C$ together with $L(0)v=nv$ for $v\in V_{(n)}$ and $Y(L(-1)v,x)=\tdif{}xY(v,x)$. The $\mbb Z$-grading of the underlying vector space of a VOA is the exact reason why the disorder operators must have integer sums when they define a VOA. On the other hand, the underlying vector space $V$ of a vertex operator superalgebra (VOSA) is a supervector space $V=V^{\bar0}\oplus V^{\bar1}$ (obeying suitably modified axioms for supervector spaces). Elements of $V^{\bar0}$ are called even, and those of $V^{\bar1}$ are called odd. The $\mbb Z/2\mbb Z$-grading leads to $\mbb Z$-valued weights of the even part $V^{\bar0}=\coprod_{n\in\mbb Z}V_{(n)}$ and $\frac12\mbb Z$-valued weights of the odd part $V^{\bar1}=\coprod_{n\in\frac12+\mbb Z}V_{(n)}$. Therefore, elements with half-integer weights are odd under the $\mbb Z/2\mbb Z$-grading. The $\frac12\mbb Z$-grading of the underlying supervector space of a VOSA is the exact reason why the disorder operators defining a VOSA must have half-integer sums.

Now we are ready to explain our main result. While the half-integer condition is powerful enough to rule out, say, $S(3)_4\to SU(3)_1$ assuming the flow admits an RG defect,\footnote{If one believes the (simple) flow exists, perform the relevant deformation in half of the space(time). What goes wrong with the picture of RG defects?} it is desirable if we could know when the sum is a half-integer and when it is an integer. The knowledge makes the constraint on RG flows stronger. Having reviewed the definitions of VO(S)As, we immediately see
\begin{equation}
    \text{VOSA}\Rightarrow\exists\mbb Z/2\mbb Z\text{-odd object}.\label{necessarycond}
\end{equation}
Namely, for the disorder operators to define an extended VOSA, there should exists a $\mbb Z/2\mbb Z$-odd operator. This gives a necessary condition for the sum to be half-integer. The contrapositive gives
\begin{equation}
    \text{No }\mbb Z/2\mbb Z\text{-odd object}\Rightarrow\text{VOA}.\label{contrapositive}
\end{equation}
Namely, if the disorder operators do not admit $\mbb Z/2\mbb Z$-odd object, the sum has to be integer
\[ h_c^\uv+h_{F(c)}^\ir\in\mbb Z. \]

How can we check whether disorder operators are $\mbb Z/2\mbb Z$-odd or not? The surviving symmetry category $\mc S_\uv$ is in general a pre-modular fusion category (pre-MFC). It is a fusion category with braiding structure and well-defined quantum dimensions. The difference between pre-MFC and MFC is that in the former, the braiding structure is not assumed to be nondegenerate. The underlying fusion category has a universal grading (possibly trivial). The universal grading is defined as follows. Let $G$ be a finite group. A $G$-grading of a fusion category $\mc C$ is a function\footnote{More precisely, the domain of the function is the set of isomorphism classes of simple objects of $\mc C$.} $\lambda:\mc C\to G$ such that $\lambda(c^*)=\lambda(c)^{-1}$ and $\lambda(c_3)=\lambda(c_1)\lambda(c_2)$ if $c_3\in c_1\otimes c_2$. The universal grading $(U(\mc C),\lambda_\mc C)$ is the universal one; for any $G$-grading $(G,\lambda)$, there exists a unique group homomorphism $\phi:U(\mc C)\to G$ such that $\lambda=\phi\lambda_\mc C$. The group $U(\mc C)$ is called the universal grading group \cite{GN06}. For example, an Ising fusion category $\ising$ with simple object ${1,x,y}$ obeying fusion products
\begin{table}[H]
\begin{center}
\begin{tabular}{c|c|c|c}
    $c\otimes c'$&$1$&$x$&$y$\\\hline
    $1$&$1$&$x$&$y$\\\hline
    $x$&$x$&$1$&$y$\\\hline
    $y$&$y$&$y$&$1\oplus x$
\end{tabular}
\end{center}
\end{table}
\hspace{-17pt}has universal $\mbb Z/2\mbb Z$-grading $U(\ising)\cong\mbb Z/2\mbb Z$ with degrees of $1,x$ are trivial, and that of $y$ is nontrivial. On the other hand, a Fibonacci fusion category $\fib$ with simple objects $1,x$ obeying fusion products
\begin{table}[H]
\begin{center}
\begin{tabular}{c|c|c}
    $c\otimes c'$&$1$&$x$\\\hline
    $1$&$1$&$x$\\\hline
    $x$&$x$&$1\oplus x$
\end{tabular}
\end{center}
\end{table}
\hspace{-17pt}only has the trivial grading, $U(\fib)\cong1$. The universal grading groups of fusion categories with small ranks can be found in the AnyonWiki \cite{anyonwiki}.\footnote{Typically, a fusion category $\mc C$ has universal $\mbb Z/2\mbb Z$-grading when $\mc C$ contains a $\mbb Z/2\mbb Z$ fusion subcategory. However, this is not a sufficient condition. For example, $\text{Rep}(S_3)$ has $\mbb Z/2\mbb Z$ fusion subcategory, while it does not admit nontrivial universal grading. When $\mc S_\uv$ is modular, it is indeed a sufficient condition thanks to a useful result, Theorem 6.3 of \cite{GN06}; a universal grading group $U(\mc C)$ of a modular category $\mc C$ is isomorphic to the group of invertible objects of $\mc C$. The theorem tells us what is $U(\mc S_\uv)$. Furthermore, when $\mc S_\uv$ is modular, $\mc S_\ir$ is also modular, and hence the product symmetry category $\mc S_\uv\boxtimes\mc S_\ir$ is so as well \cite{M02}. Thus, a universal grading group of the product symmetry category $\mc S_\uv\boxtimes\mc S_\ir$ is given by $U(\mc S_\uv)\times U(\mc S_\uv)$ when $\mc S_\uv$ is modular. (Note that since $\mc S_\ir$ has the same fusion rules as $\mc S_\uv$, we have $U(\mc S_\ir)\cong U(\mc S_\uv)$.)} Here, we can prove a\\

\textbf{Lemma.} \textit{Let $\mc C,\mc D$ be fusion categories. We have $U(\mc C\boxtimes\mc D)\cong U(\mc C)\times U(\mc D)$.}\\

\textit{Proof.} By assumption, we have $\forall c_1,c_2\in\mc C,\forall d_1,d_2\in\mc D,\lambda_\mc C(c_1^*)=\lambda_\mc C(c_1)^{-1},\lambda_\mc C(c_1\otimes c_2)=\lambda_\mc C(c_1)\lambda_\mc C(c_2)$ and $\lambda_\mc D(d_1^*)=\lambda_\mc D(d_1)^{-1},\lambda_\mc D(d_1\otimes d_2)=\lambda_\mc D(d_1)\lambda_\mc D(d_2)$. Objects of $\mc C\boxtimes\mc D$ are given by $c\boxtimes d$ with $c\in\mc C,d\in\mc D$. They obey fusion
\[ (c_1\boxtimes d_1)\otimes(c_2\boxtimes d_2)\cong(c_1\otimes c_2)\boxtimes(d_1\otimes d_2). \]
Thus, we can define a function $\lambda_{\mc C\boxtimes\mc D}:\mc C\boxtimes\mc D\to U(\mc C)\times U(\mc D)$ to a group $U(\mc C)\times U(\mc D)$ by $\lambda_{\mc C\boxtimes\mc D}(c\boxtimes d):=\lambda_\mc C(c)\lambda_\mc D(d)$. The function obeys
\[ \lambda_{\mc C\boxtimes\mc D}((c\boxtimes d)^*)=\lambda_{\mc C\boxtimes\mc D}(c^*\boxtimes d^*)=\lambda_\mc C(c)^{-1}\lambda_\mc D(d)^{-1} \]
and
\begin{align*}
    \lambda_{\mc C\boxtimes\mc D}((c_1\boxtimes d_1)\otimes(c_2\boxtimes d_2))&=\lambda_{\mc C\boxtimes\mc D}((c_1\otimes c_2)\boxtimes(d_1\otimes d_2))\\
    &=\lambda_\mc C(c_1\otimes c_2)\lambda_\mc D(d_1\otimes d_2)\\
    &=\lambda_\mc C(c_1)\lambda_\mc C(c_2)\lambda_\mc D(d_1)\lambda_\mc D(d_2).
\end{align*}
The function is universal; $\forall\lambda=(\lambda_1,\lambda_2):\mc C\boxtimes\mc D\to G_1\times G_2,\exists!\phi=(\phi_1,\phi_2):U(\mc C)\times U(\mc D)\to G_1\times G_2$ such that
\[ \lambda=(\lambda_1,\lambda_2)=(\phi_1\lambda_\mc C,\phi_2\lambda_\mc D)=\phi\lambda_{\mc C\boxtimes\mc D} \]
by the universality of $(U(\mc C),\lambda_\mc C),(U(\mc D),\lambda_\mc D)$. $\square$\\

With the universal grading, it is not difficult to check whether a given disorder operator is $\mbb Z/2\mbb Z$-odd or not; the disorder operator at the end of $cF(c)$ is $\mbb Z/2\mbb Z$-odd if the surviving symmetry category $\mc S_\uv$ has $\mbb Z/2\mbb Z\subset U(\mc S_\uv)$ and if $c\in\mc S_\uv$ has odd degree under the $\mbb Z/2\mbb Z$. Hence, the non-invertible symmetry object $y\in\ising$ could give a disorder operator defining VOSA, while $\fib$ cannot define VOSA because the fusion category does not admit universal $\mbb Z/2\mbb Z$-grading. Therefore, as a corollary of (\ref{contrapositive}), we get
\begin{equation}
    \mbb Z/2\mbb Z\notsubset U(\mc S_\uv)\Rightarrow h_c^\uv+h_{F(c)}^\ir\in\mbb Z\quad(c\in\mc S_\uv).\label{integercond}
\end{equation}
This implies the image $\mc S_\ir$ of $\mc S_\uv$ under the monoidal functor has the \textit{opposite} braiding structure.\footnote{This further implies the product symmetry category is the Drinfeld center of the surviving symmetry category $\mc S_\uv$
\[ \mc S_\uv\boxtimes\mc S_\ir\simeq\mc S_\uv\boxtimes\widetilde{\mc S_\uv}\simeq\mc Z(\mc S_\uv) \]
when $\mc S_\uv$ is modular \cite{M01,DGNO09}. Since the Drinfeld center admits Lagrangian algebra \cite{DMNO10,M12}, such RG flows are conjecturally defined \cite{KKWitt} by the category of vector spaces $\vect\in[\mc S_\uv\boxtimes\mc S_\ir]$.} Our main observation should become clear through examples. One can see the pattern if we modify the Table 1 in \cite{KKWitt}:
\begin{table}[H]
\begin{center}
\makebox[1 \textwidth][c]{       
\resizebox{1.2 \textwidth}{!}{\begin{tabular}{c|c|c|c|c|c}
Rational RG flow&Surviving symmetry $\mc S_\uv$&$\mbb Z/2\mbb Z\subset U(\mc S_\uv)$?&$h_c^\uv+h_{F(c)}^\ir$&$\frac12\mathbb Z$-spin&Extended algebra\\\hline
$M(5,4)\to M(4,3)$&$\ising$ MFC&Yes&$(0,2,\frac12)$&$\mbb Z/2\mbb Z$-odd object KW&VOSA\\
$M(6,5)\to M(5,4)$&$\fib\boxtimes\vecG_{\mbb Z/2\mbb Z}^1$ pre-MFC&Yes&$(0,2,\frac92,\frac12)$&$\mbb Z/2\mbb Z$-odd objects&VOSA\\
$M(7,6)\to M(6,5)$&$su(2)_4$ MFC&Yes&$(0,0,\frac12,\frac92,2)$&$\mbb Z/2\mbb Z$-odd objects&VOSA\\
$M(8,7)\to M(7,6)$&$\svec\boxtimes psu(2)_5$ pre-MFC&Yes&$(0,8,2,\frac{25}2,\frac12,\frac92)$&$\mbb Z/2\mbb Z$-odd objects&VOSA\\
$M(3,5)\to M(2,5)$&$\fib$ MFC&No&$(0,0)$&&VOA\\
$M(15,4)\to M(9,4)$&$\ising$ MFC&Yes&$(0,10,\frac72)$&$\mbb Z/2\mbb Z$-odd object KW&VOSA\\
$\wf{su}(3)_2\to\wf{su}(3)_1$&$\vecG_{\mbb Z/3\mbb Z}^1$ MFC&No&$(0,1,1)$&&VOA\\
$\wf{(e_7)}_3\to\wf{(e_7)}_1$&$\vecG_{\mbb Z/2\mbb Z}^{-1}$ MFC&Yes&$(0,3)$&None&VOA\\
$(A_{10},E_6)\to M(4,3)$&$\ising$ MFC&Yes&$(0,4,2)$&None&VOA
\end{tabular}.}}
\end{center}
\caption{Summary of examples from \cite{KKWitt}}
\end{table}\hspace{-17pt}On the other hand, it is unclear when the sum becomes half-integer. One can see the same $\mc S_\uv\simeq\ising$ (as fusion categories) with $\mbb Z/2\mbb Z$ universal grading obeys the integer condition in one example and half-integer condition in another. The symmetry categories may not be sufficient to tell when the sum is a half-integer. 

Beside the half-integer condition, we also employ the spin constraints found in \cite{KKSUSY}. The constraint works as follows. In (\ref{defectHilbertspace}), we saw the defect Hilbert space $\mc H_c^\uv$ has physical interpretation. The expression tells us that states (or operators via state-operator correspondence) have specific spins $h_i-h_j\mod1$ when $(SM_cS^\dagger)_{i,j}\in\mbb N$ is nonzero. The set $S_c^\uv$ of spins is called the spin content. Now let us deform the UV theory with relevant operators. If they are space(time) scalars, the rotation symmetry is preserved. Hence, spin quantum numbers are conserved. If the RG flow reaches another RCFT, we get another defect Hilbert space $\mc H_{F(c)}^\ir$ for a surviving symmetry operator $c$. It again has specific spins and we get another spin content $S_{F(c)}^\ir$. Since spin quantum numbers are conserved, they should basically match. More precisely, heavy operators might be lifted along the RG flow, and $\mc H_{F(c)}^\ir$ might no longer have them. Therefore, the spin content in IR should be a subset of the spin content in UV:
\begin{equation}
    S_{F(c)}^\ir\subset S_c^\uv\quad(c\in\mc S_\uv).\label{spinconstraint}
\end{equation}

\section{Examples}
In this section, we solve some RG flows. In order to see the power of our result (\ref{integercond}), we limit our examples to flows with surviving symmetry $\mc S_\uv$ such that $\mbb Z/2\mbb Z\notsubset U(\mc S_\uv)$. Throughout the section, we assume massless RG flows admit RG defects.\footnote{Note that in all the examples, the conjectured ``$h$-theorem'' \cite{KK22II,KK22free,KKARG,KKWZW,KK25}
\[ h_{F(c)}^\ir\le h_c^\uv\quad(c\in\mc S_\uv) \]
holds.}

We take diagonal unitary minimal models as our UV theories. (Thus, they have central charges $0<c_\uv<1$.) As we reviewed above, their symmetry categories are MFCs. We deform the UV theories with relevant operator(s). The deformation preserves a pre-modular fusion subcategory $\mc S_\uv$. If the flow is massless, the IR theories have scale symmetry. Since our RG flows preserve unitarity, the scale symmetry enhances to the full conformal symmetry by the Zamolodchikov-Polchinski theorem \cite{cthm,P87}. The central charges of the IR CFTs are in the range $0<c_\ir<c_\uv<1$ due to the $c$-theorem \cite{cthm}. Such CFTs have been classified \cite{KL02}, and all of them are rational. Therefore, we can apply our result. If the flow is massive, we can employ another mathematical fact; the gapped IR phases are described by (left) $\mc S_\uv$-module categories \cite{TW19,HLS21}. For example, for $\mc S_\uv\simeq\fib$, indecomposable $\fib$-module categories have been classified in \cite{BD11,KK23GSD,KK23preMFC,KK23rank5}. The result is that the only indecomposable $\fib$-module category is $\fib$ itself. Therefore, we learn the ground state degeneracy (GSD) in the massive flows with $\fib$ should be a multiple of two. We also conjecture signs of the relevant couplings employing the Cardy's method \cite{C17}.

\subsection{$(E_8,A_{30})+|\phi_{\frac{29}{31}}|^2\to(A_{28},E_8)$}
The UV RCFT has rank 30 modular symmetry category
\[ \fib\boxtimes psu(2)_{29}. \]
The theory has 14 relevant primaries. We take the least relevant one with the conformal dimension $h=\frac{29}{31}$ as our relevant operator. It obeys the operator product expansion (OPE)
\[ \phi_{\frac{29}{31}}\times\phi_{\frac{29}{31}}\sim1+\phi_{\frac{29}{31}}+\phi_{\frac{118}{31}}, \]
and no other relevant operators are turned on. Since the relevant operator has Kac index $(1,3)$, the flow was studied in \cite{R91}. It was conjectured that, in the massless scenario, the IR theory is the $(A_{28},E_8)$ minimal model based mainly on the conservation of ADE algebras. Let us reanalyze the scenario from the viewpoint of symmetry.

The relevant deformation preserves the Fibonacci MFC
\[ \mc S_\uv\simeq\fib, \]
where the Fibonacci object has conformal dimension $\frac{47}5$. The symmetry category has central charge
\[ c(\mc S_\uv)=\frac{14}5\mod8. \]
As we reviewed above, if the flow is massless, the IR theory is an RCFT with smaller central charge and with Fibonacci symmetry category. In general, we have to search for consistent MFCs with the smallest global dimensions \cite{KK21,KK22II,KK22free}, however, in our given problem, we can employ the commutativity of relevant deformation and discrete gauging \cite{KK22II}. In the $A$-type minimal models, RG flows $M(m+1,m)+\phi_{1,3}\to M(m,m-1)$ are known (for $m>3$). Our UV theory is obtained from the $M(31,30)$ by gauging $1\oplus c_{26}\oplus c_{84}\oplus c_{203}$ where we labeled nontrivial symmetry objects with their conformal dimensions. Since the four simple objects act trivially on our relevant operator $\phi_{1,3}$, the relevant deformation and the discrete gauging commute; two results obtained by (1) first deforming $M(31,30)$ with $\phi_{1,3}$ and then gauging the object (which survives the deformation), or (2) first gauging the object in $M(31,30)$ and then deforming with $\phi_{1,3}$, are the same. The first path (1) gives gauging of $M(30,29)$, which is the $(A_{28},E_8)$ theory. Therefore, we learn the second path (2) also gives the theory $(A_{28},E_8)$.

In fact, we find the $(A_{28},E_8)$ satisfies all constraints. The spin constraint \cite{KKSUSY} is satisfied because
\[ S_\fib^{(E_8,A_{30})}=\{0,\pm\frac25\}=S_\fib^{(A_{28},E_8)}\quad\mod1. \]
If we assume the flow admits an RG defect, it should satisfy our integer condition (\ref{integercond}) because $\fib$ does not admit a universal $\mbb Z/2\mbb Z$-grading. Since the Fibonacci object of $(A_{28},E_8)$ has conformal dimension $\frac{43}5$, the integer condition is beautifully satisfied
\[ \frac{47}5+\frac{43}5\in\mbb Z. \]
Finally, we should explain why modular $psu(2)_{27}$ emerges in IR. First, since symmetry categories of RCFTs are modular, the IR theory should have modular symmetry category containing the surviving $\fib$. Naively, one may think no emergent symmetries are needed because $\fib$ is already modular. However, consistency with the $c$-theorem \cite{KK21} requires them. The image MFC $\mc S_\ir\simeq\fib$ with $h_\fib=\frac35\mod1$ has central charge
\[ c(\fib)=-\frac{14}5\mod8. \]
In unitary theories, this means $c=\frac{26}5,\frac{66}5,\dots$, which \textit{cannot} be smaller than $c_\uv=\frac{154}{155}$. The consistency demands an emergent symmetry. This is a mathematical reason why there should exist an emergent symmetry in the massless scenario. Second, for emergent symmetries and $\fib$ to form an MFC, emergent symmetries themselves should form an MFC $\mc B$. For the IR symmetry $\fib\boxtimes\mc B$ to be consistent with the $c$-theorem, it should obey
\[ 0<-\frac{14}5+c(\mc B)<c_\uv=\frac{154}{155}\mod8. \]
In fact, since discrete gaugings do not change central charges,\footnote{Physical proof: One way to define central charges is through OPE of energy-momentum tensors. By definition, energy-momentum tensors commute with (internal) symmetry operators. Therefore, gauging (discrete) symmetries does not change OPE, and hence central charges remain intact.} the emergent MFC should obey
\[ -\frac{14}5+c(\mc B)=c_{M(30,29)}=\frac{144}{145}\mod8. \]
Going through the list of MFCs \cite{GK94,RSW07,BNRW15,LPR20,VS22,NRW23} up to rank 12, we find \textit{none} of them satisfy the condition. (Since the RG flow preserves unitarity, we can limit our search to unitary MFCs.) This shows the emergent MFC $\mc B$ should have rank no smaller than 13. The result is also consistent with the known scenario; the $(A_{28},E_8)$ has rank 28 modular symmetry category
\[ \fib\boxtimes psu(2)_{27}. \]
The second factor is a rank 14 MFC. We summarize our analysis as the\\

\textbf{Conjecture.} \textit{The unitary modular fusion category with central charge $\frac{110}{29}\mod8$ and with the minimum global dimension\footnote{The global dimension $D^2$ of an MFC $\mc C$ is defined as
\[ D^2:=\sum_{i=1}^{\rank(\mc C)}d_i^2. \]
The quantum dimension of a Verlinde line $c_i\in\mc C$ is given by the ratio of modular $S$-matrices:
\[ d_i=\frac{S_{i1}}{S_{11}}, \]
where 1 denotes the identity operator.} is given by the $psu(2)_{27}=\{1,c_2,c_3,\dots,c_{14}\}$ modular fusion category with quantum
\[ (d_2,d_3,\dots,d_{14})=(\frac{\sin\frac{3\pi}{29}}{\sin\frac\pi{29}},\frac{\sin\frac{5\pi}{29}}{\sin\frac\pi{29}},\dots,\frac{\sin\frac{27\pi}{29}}{\sin\frac\pi{29}}) \]
and conformal dimensions
\[ (h_2,h_3,\dots,h_{14})=(\frac2{29},\frac6{29},\frac{12}{29},\frac{20}{29},\frac1{29},\frac{13}{29},\frac{27}{29},\frac{14}{29},\frac3{29},\frac{23}{29},\frac{16}{29},\frac{11}{29},\frac8{29})\mod1. \]}\\

Before we move to the next example, let us briefly mention on the massive scenario. In this case, the IR topological quantum field theory (TQFT) is described by a $\fib$-module category. Since the only indecomposable $\fib$-module category is itself (i.e., regular module category), the GSD should be a multiple of two. We can further give conjectures on signs of the relevant coupling $\lambda_{1,3}$ with the Cardy's method \cite{C17}. In the method, with the Lagrangian coupling $\delta S\ni\lambda_{1,3}\int|\phi_{\frac{29}{31}}|^2$, we try to minimize the energy
\[ E_a=\frac{\pi\cdot154/155}{24(2\tau_1)^2}-\frac{S_{ja}}{S_{1a}}\frac{\lambda_{1,3}}{(2\tau_a)^{58/31}} \]
for a Cardy state $|a\ra$ with variational parameter $\tau_a$. Here, $j$ is the index corresponding to the relevant operator $\phi_{\frac{29}{31}}$. The ratio of the modular $S$-matrices takes the largest value ($\approx2.959$) for Cardy states with conformal dimensions $0,\frac{47}5$, and the smallest value ($\approx-0.990$) for those with $\frac{99}{31},\frac{92}{155}$. Therefore, for positive Lagrangian couplings, the first two Cardy states give (approximate) vacuum, and for negative Lagrangian couplings, the second two Cardy states give twofold degenerate vacua. Since the first Cardy states include the identity, this signals the massless flow. To summarize, we arrive the conjecture
\begin{equation}
    (E_8,A_{30})+\lambda_{1,3}|\phi_{\frac{29}{31}}|^2=\begin{cases}(A_{28},E_8)&(\lambda_{1,3}>0),\\\text{TQFT w/ GSD}=2&(\lambda_{1,3}<0).\end{cases}\label{E8A30result}
\end{equation}

\subsection{$(A_{28},E_8)+|\phi_\frac{20}{29}|^2+|\phi_\frac1{29}|^2+|\phi_\frac{14}{29}|^2+|\phi_\frac3{29}|^2+|\phi_\frac8{29}|^2\to(A_4,D_4)$}
The previous example was known, and in some sense, it was not that interesting. Let us apply our method to unknown RG flows. In this example, we take the $(A_{28},E_8)$ minimal model as our UV theory. The UV RCFT has rank 28 modular symmetry category
\[ \fib\boxtimes psu(2)_{27}. \]
The theory has 13 relevant primaries. Those with conformal dimensions $\frac{20}{29},\frac1{29},\frac{14}{29},\frac3{29},\frac8{29}$ preserve $\fib$, and the other eight preserve only the identity symmetry. In order to get nontrivial IR theories, we take our relevant operator(s) from the first class. Namely, we turn on all the relevant operators. This is inevitable because one relevant operator turns on the others. One way to see this is to see the OPEs. For example, we have
\begin{align*}
    \phi_{\frac{20}{29}}\times\phi_{\frac{20}{29}}&\sim1+\phi_{\frac{31}{29}}+\phi_{\frac{122}{29}}+\phi_{\frac{99}{29}}+\phi_{\frac{20}{29}}+\phi_{\frac1{29}}+\phi_{\frac{42}{29}}+\phi_{\frac{85}{29}}+\phi_{\frac{14}{29}},\\
    \phi_{\frac{20}{29}}\times\phi_{\frac{14}{29}}&\sim\phi_{\frac{20}{29}}+\phi_{\frac1{29}}+\phi_{\frac{42}{29}}+\phi_{\frac{85}{29}}+\phi_{\frac{14}{29}}+\phi_{\frac3{29}}+\phi_{\frac{52}{29}}+\phi_{\frac{161}{29}}+\phi_{\frac{69}{29}},\\
    \phi_{\frac{20}{29}}\times\phi_{\frac3{29}}&\sim\phi_{\frac1{29}}+\phi_{\frac{42}{29}}+\phi_{\frac{85}{29}}+\phi_{\frac{14}{29}}+\phi_{\frac3{29}}+\phi_{\frac{52}{29}}+\phi_{\frac{161}{29}}+\phi_{\frac{69}{29}}+\phi_{\frac8{29}}.
\end{align*}
Let us study the RG flow\footnote{One way to obtain $(A_{28},E_8)$ model is to gauge non-invertible symmetry of $M(30,29)$. Since the symmetries to be gauged in $M(30,29)$ act nontrivially on relevant operators forming primaries in $(A_{28},E_8)$, the commutativity cannot be used in this example.} from the viewpoint of symmetry.

The relevant deformation preserves the Fibonacci MFC
\[ \mc S_\uv\simeq\fib, \]
where the Fibonacci object has conformal dimension $\frac{43}5$. The symmetry category has central charge
\[ c(\mc S_\uv)=-\frac{14}5\mod8. \]
We perform a case analysis. If the RG flow is massless, the IR theory should be an RCFT with central charge smaller than $c_\uv=\frac{144}{145}$ and with Fibonacci symmetry objects. There are only five candidates: $(D_6,A_{10}),(A_8,D_6),(A_4,D_4),M(6,5),M(5,4)$. All of them satisfy the spin constraint
\[ S_\fib^{(A_{28},E_8)}=\{0,\pm\frac25\}=S_\fib^\text{five candidates}\quad\mod1. \]
Now, let us assume the flow admits an RG defect. Then, it should obey the integer condition (\ref{integercond}). We find only three $(A_8,D_6),(A_4,D_4),M(6,5)$ satisfy the conditions because their Fibonacci objects have conformal dimensions $(\frac{17}5,\frac{17}5)$, $\frac25$, and $\frac75$, respectively. (The first $(A_8,D_6)$ has two Fibonacci objects because its MFC has a subcategory $\fib\boxtimes\fib$.) Which RCFTs are realized in IR? Although we cannot rule out either of them, the observation \cite{KK22free} favors the second. An MFC with global dimension $D^2$ contributes free energy by
\[ F\ni T\ln D, \]
where $T$ is the temperature defined as the length $1/T$ of the Euclidean time compactified to a circle. Here, $-\ln D$ is the topological entanglement entropy \cite{KP05,LW05}. Since physical processes minimize free energies, if there are multiple consistent MFCs, an MFC with the smallest global dimension is energetically favored. Thus, our search for IR theory reduces to this question: Which MFC has the smallest global dimension? They have global dimensions
\[ \hspace{-35pt}D^2_{(A_8,D_6)}=\frac9{4\sin^2\frac\pi9}\frac{15+5\sqrt5}2\approx251.782,\quad D^2_{(A_4,D_4)}=3\times\frac{5+\sqrt5}2\approx10.854,\quad D^2_{M(6,5)}=12\times\frac{5+\sqrt5}2\approx43.416. \]
Therefore, we see that $(A_4,D_4)$ is favored.\footnote{There are five relevant couplings. We cannot rule out the possibilities that their fine-tunings would lead to the other two theories. Our claim is that for \textit{generic} relevant couplings, $(A_4,D_4)$ is energetically favored.}

Let us also comment on emergent symmetry. The scenario says there are emergent symmetries. We should explain why they appear. The image $\mc S_\ir$ of $\mc S_\uv$ under the monoidal functor giving the massless RG flow has a Fibonacci object with conformal dimension $\frac25$ mod 1 and central charge
\[ c(\mc S_\ir)=\frac{14}5\mod8. \]
In unitary theories, this means $c=\frac{14}5,\frac{54}5,\dots$. In the absence of emergent symmetries, the central charge \textit{cannot} be smaller than $c_\uv=\frac{144}{145}$, which is inconsistent with the $c$-theorem. The consistency with the theorem requires an emergent symmetry. We repeat the same search as in the previous example. Since $\fib$ is modular, emergent symmetries should form an MFC $\mc B$ with central charge
\[ 0<\frac{14}5+c(\mc B)<c_\uv=\frac{144}{145}\mod8. \]
No rank two MFC $\mc B$ satisfies these conditions. At rank three, we find $\mc B\simeq\vecG_{\mbb Z/3\mbb Z}^1$ with $h_{\mbb Z/3\mbb Z}=\frac23$ and $\mc B\simeq\ising$ with $h_{KW}=\frac{11}{16}$ (mod 1) satisfy the conditions. However, there is no RCFT for the latter because it gives $c=\frac3{10}$. The minimal possibility $\mc B\simeq\vecG_{\mbb Z/3\mbb Z}^1$ gives the exact symmetry category of $(A_4,D_4)$. This analysis\footnote{One can continue the search for consistent MFCs $\mc B$. At rank four, no MFC is consistent. At rank five, one finds two MFCs are consistent: two $\mc B\simeq su(2)_4$ with conformal dimensions $(h_2,h_3,h_4)=(\frac18,\frac58,\frac23)$ and $(\frac38,\frac78,\frac23)$. The former gives the tricritical Ising model $M(6,5)$.} also supports our proposal $(A_{28},E_8)\to(A_4,D_4)$. This concludes our analysis for the massless case. For the massive case, the IR theory should be a TQFT with GSD a multiple of two. To summarize, we conjecture
\begin{equation}
    (A_{28},E_8)+|\phi_\frac{20}{29}|^2+|\phi_\frac1{29}|^2+|\phi_\frac{14}{29}|^2+|\phi_\frac3{29}|^2+|\phi_\frac8{29}|^2=\begin{cases}(A_4,D_4)&(\text{massless}),\\\text{TQFT w/ GSD}\in2\mbb N^\times&(\text{massive}).\end{cases}\label{A28E8result}
\end{equation}

\subsection{$(D_6,A_{10})+|\phi_{\frac9{11}}|^2\to(A_8,D_6)$}
Let us also apply our method to $D$-type models. As our third example, we pick the $(D_6,A_{10})$ minimal model as our UV theory. The theory has rank 20 modular symmetry category
\[ psu(2)_9\boxtimes\fib\boxtimes\fib, \]
where the two Fibonacci objects both have conformal dimensions $h_\fib=\frac{23}5$. We deform the theory with the least relevant operator with conformal dimension $\frac9{11}$. It does not turn on other relevant operators because it obeys the OPE
\[ \phi_{\frac9{11}}\times\phi_{\frac9{11}}\sim1+\phi_{\frac9{11}}+\phi_{\frac{38}{11}}. \]
Since it has Kac index $(1,3)$, this RG flow was also studied in \cite{R91}. In the massless case, the IR theory was conjectured to be $(A_8,D_6)$. Let us reanalyze the flow from the viewpoint of symmetry.

The relevant deformation preserves the MFC
\[ \mc S_\uv\simeq\fib\boxtimes\fib. \]
It has central charge
\[ c(\mc S_\uv)=-\frac{28}5\mod8. \]
We perform a case analysis. If the flow is massless, the IR theory is an RCFT with central charge smaller than $c_\uv=\frac{52}{55}$ which contains $\fib\boxtimes\fib$ as a subcategory. The only candidate is the $(A_8,D_6)$ minimal model. Indeed, the conjecture satisfies all the constraints. The spin constraint is satisfied as
\begin{align*}
    S_x^{(D_6,A_{10})}=&\{0,\pm\frac25\}=S_X^{(A_8,D_6)}\mod1,\\
    S_{x'}^{(D_6,A_{10})}=&\{0,\pm\frac25\}=S_{X'}^{(A_8,D_6)}\mod1,\\
    S_{xx'}^{(D_6,A_{10})}=&\{0,\pm\frac15,\pm\frac25\}=S_{XX'}^{(A_8,D_6)}\mod1,
\end{align*}
where we denote Fibonacci objects in the UV theory by $x,x'$, and their images in the IR theory by $X,X'$. Now, let us assume our flow admits an RG defect. Then, it should obey the integer condition (\ref{integercond}) because $\fib\boxtimes\fib$ does not admit universal $\mbb Z/2\mbb Z$-grading. We learn the IR theory should have
\[ \mc S_\ir\simeq\fib\boxtimes\fib \]
with conformal dimensions $h_\fib=\frac25\mod1$. The condition is beautifully satisfied because
\begin{align*}
    h_x^{(D_6,A_{10})}+h_X^{(A_8,D_6)}&=\frac{23}5+\frac{17}5=8,\\
    h_{x'}^{(D_6,A_{10})}+h_{X'}^{(A_8,D_6)}&=\frac{23}5+\frac{17}5=8,\\
    h_{xx'}^{(D_6,A_{10})}+h_{XX'}^{(A_8,D_6)}&=\frac65+\frac45=2.
\end{align*}
Such an MFC has central charge
\[ c(\mc S_\ir)=\frac{28}5\mod8. \]
This \textit{cannot} be smaller than $c_\uv$, and the consistency with the $c$-theorem requires an emergent symmetry. Since the IR theory is rational, emergent symmetries should form an MFC $\mc B$ with
\[ 0<\frac{28}5+c(\mc B)<c_\uv=\frac{52}{55}. \]
Going through the list of MFCs, we find no consistent MFCs at rank two and three.\footnote{We do find MFCs giving central charges in the range, however, there are no RCFTs with the central charges.} At rank four, we find only one consistent MFC
\[ \mc B\simeq psu(2)_7 \]
with conformal dimensions $(h_2,h_3,h_4)=(\frac13,\frac29,\frac23)\mod1$. It has
\[ c(psu(2)_7)=\frac{10}3\mod8, \]
and gives
\[ c(\mc S_\ir\boxtimes psu(2)_7)=\frac{28}5+\frac{10}3=\frac{134}{15}=\frac{14}{15}\mod8. \]
The product MFC $\fib\boxtimes\fib\boxtimes psu(2)_7$ is the exact symmetry of $(A_8,D_6)$, and the central charge matches that of the theory. Our symmetry analysis supports the conjecture in \cite{R91}. If the flow is massive, the IR theory is described by a $\fib\boxtimes\fib$-module category. Since $\fib\subset\fib\boxtimes\fib$, we know the module category should have an even rank. However, to the best of our knowledge, there is no complete classification of indecomposable $\fib\boxtimes\fib$-module categories. Therefore, we directly resort to the Cardy's method. In the method, we try to minimize the energy
\[ E_a=\frac{\pi\cdot52/55}{24(2\tau_1)^2}-\frac{S_{ja}}{S_{1a}}\frac{\lambda_{1,3}}{(2\tau_a)^{18/11}}, \]
where $j$ is the index corresponding to the primary $\phi_{\frac9{11}}$. The ratio of modular $S$-matrices takes the largest value $(\approx2.683)$ for Cardy states with conformal dimensions $0,\frac{23}5,\frac{23}5,\frac65$, and the smallest value $(\approx-0.919)$ for those with $\frac{38}{11},\frac3{55},\frac3{55},\frac{36}{55}$. The computation signals massless flow for $\lambda_{1,3}>0$, and massive flow with GSD $=4$ for $\lambda_{1,3}<0$:
\begin{equation}
    (D_6,A_{10})+\lambda_{1,3}|\phi_{\frac9{11}}|^2=\begin{cases}(A_8,D_6)&(\lambda_{1,3}>0),\\\text{TQFT w/ GSD}=4&(\lambda_{1,3}<0).\end{cases}\label{D6A10result}
\end{equation}

\subsection{$(D_4,A_6)+|\phi_{\frac17}|^2+|\phi_{\frac57}|^2\to(A_4,D_4)$}
As the last example, we pick the $(D_4,A_6)$ minimal model as our UV theory. It has rank nine modular symmetry category
\[ \vecG_{\mbb Z/3\mbb Z}^1\boxtimes psu(2)_5. \]
The theory has six relevant operators with conformal dimensions $\frac17,\frac57,\frac{10}{21},\frac{10}{21},\frac1{21},\frac1{21}$. The last four only preserve the identity symmetry, while the first two preserve $\vecG_{\mbb Z/3\mbb Z}^1$. In order to get nontrivial RG flow, we choose our relevant operators from the second class. Since they obey the OPE
\begin{align*}
    \phi_{\frac17}\times\phi_{\frac17}&\sim1+\phi_{\frac57},\\
    \phi_{\frac57}\times\phi_{\frac57}&\sim1+\phi_{\frac17}+\phi_{\frac57},
\end{align*}
turning on one turns on the other. The least relevant primary $\phi_{\frac57}$ has Kac index, and \cite{R91} studied massless RG flow triggered by the primary. However, as we saw now, one cannot avoid turning on both relevant operators, and strictly speaking, our RG flow goes beyond the one in the paper.\footnote{For example, in equation (16) of the paper, the OPE coefficients $C_{(13),j}^j$ were considered, but one also needs to study those for $\phi_{\frac17}$ with Kac index $(1,2)$.} Let us analyze the flow from the viewpoint of symmetry.

As we mentioned above, the relevant deformation preserves the $\mbb Z/3\mbb Z$ MFC
\[ \mc S_\uv\simeq\vecG_{\mbb Z/3\mbb Z}^1. \]
The $\mbb Z/3\mbb Z$ objects have conformal dimensions $\frac43$, and the MFC has central charge
\[ c(\mc S_\uv)=2\mod8. \]
If one knows the list of (R)CFTs with smaller central charges, one immediately finds the only possible IR theory in massless case is the three-state Potts model $(A_4,D_4)$ because it is the only lower theory with $\vecG_{\mbb Z/3\mbb Z}^1$ MFC. Here, in order to see how our method works, let us pretend we do not know the list, and let us try to \textit{discover} the theory.

We perform case analysis. In the massless case, assuming an RG defect exists, we know
\[ \mc S_\ir\simeq\vecG_{\mbb Z/3\mbb Z}^1 \]
with conformal dimensions $h_{\mbb Z/3\mbb Z}=\frac23\mod1$. Such an MFC has central charge
\[ c(\mc S_\ir)=-2\mod8. \]
In unitary theories, this means $c=6,14,\dots$, which \textit{cannot} be smaller than $c_\uv=\frac67$. Therefore, for an IR theory to be massless, there should exist an emergent symmetry. Since $\mc S_\ir\simeq\vecG_{\mbb Z/3\mbb Z}^1$ is already modular, for an IR theory to have modular symmetry category, emergent symmetries should form an MFC $\mc B$ such that
\[ 0<-2+c(\mc B)<c_\uv=\frac67\mod8. \]
Going through the list of MFCs, we find there is one consistent MFC at rank two\footnote{If one continues the search for consistent MFCs for higher ranks, one finds none at rank three (or higher). Since our consistent MFC $\mc B\simeq\fib$ has global dimension $D^2=\frac{5+\sqrt5}2\approx3.618$, we can stop our search here. MFCs with higher ranks have larger global dimensions, and they are energetically disfavored.}
\[ \mc B\simeq\fib \]
with conformal dimension $h_\fib=\frac25\mod1$. The MFC has central charge $c(\fib)=\frac{14}5$, and gives
\[ c(\mc S_\ir\boxtimes\fib)=-2+\frac{14}5=\frac45\mod8. \]
The modular symmetry category $\vecG_{\mbb Z/3\mbb Z}^1\boxtimes\fib$ is the exact symmetry of the three-state Potts model. This scenario is also consistent with the spin constraint
\[ S_{x,x^2}^{(D_4,A_6)}=\{0,\pm\frac13\}=S_{X,X^2}^{(A_4,D_4)}\mod1, \]
where we denote $\mbb Z/3\mbb Z$ objects in UV by $x,x^2$, and their images in IR by $X,X^2$. We have $h_X=\frac23=h_{X^2}$. This concludes our analysis for the massless case. For the massive case, we know the IR TQFT is described by a $\vecG_{\mbb Z/3\mbb Z}^1$-module category.\footnote{The classification of indecomposable $\vecG_{\mbb Z/3\mbb Z}^1$-module categories have been performed in \cite{KK23preMFC,KK24rank9} employing the method developed in \cite{KK23GSD,KK23preMFC,KK23rank5,KKH24,KK24rank9}. The set of indecomposable $\mc C$-module categories stands in bijection with another set of Lagrangian algebras in $\mc Z(\mc C)$ for a fusion category $\mc C$ \cite{DMNO10}. In \cite{KK24rank9}, it was found that $\mc Z(\vecG_{\mbb Z/3\mbb Z}^1)$ has two Lagrangian algebras, thus two indecomposable $\vecG_{\mbb Z/3\mbb Z}^1$-module categories. The indecomposable module categories were not identified in the paper, but it is not difficult to identify them. First, any fusion category has itself as an indecomposable module category called regular module category. Second, since $\mbb Z/3\mbb Z$ is anomaly-free, we know there exists rank one module category. Therefore, the two indecomposable $\vecG_{\mbb Z/3\mbb Z}^1$-module categories have ranks three and one.} Since the surviving symmetry is an anomaly-free $\mbb Z/3\mbb Z$ symmetry, both one- and three-dimensional modules (i.e., representations) exist:
\begin{equation}
    (D_4,A_6)+|\phi_{\frac17}|^2+|\phi_{\frac57}|^2=\begin{cases}(A_4,D_4)&(\text{massless}),\\\text{TQFT described by a }\vecG_{\mbb Z/3\mbb Z}^1\text{-module category}&(\text{massive}).\end{cases}\label{D4A6result}
\end{equation}

\section*{Acknowledgment}
We acknowledge support of the Department of Atomic Energy, Government of India, under project no. RTI4019.

\appendix
\setcounter{section}{0}
\renewcommand{\thesection}{\Alph{section}}
\setcounter{equation}{0}
\renewcommand{\theequation}{\Alph{section}.\arabic{equation}}

\end{document}